\documentclass[amsfonts, amssymb, amsmath, showkeys, nofootinbib, preprint,floatfix,aps, prx, superscriptaddress]{revtex4-2}

\usepackage[english]{babel}
\usepackage[colorinlistoftodos, color=green!40,prependcaption]{todonotes}
\usepackage{graphicx}
\usepackage{dcolumn}
\usepackage{bm}
\usepackage{float}
\usepackage{gensymb}
\usepackage{xcolor}
\usepackage{amsmath}
\usepackage{amssymb}
\usepackage{units}
\usepackage{placeins}
\usepackage{ulem}
\usepackage{hyperref}
\usepackage{bbm}
\usepackage{multirow}
\usepackage{amsfonts}
\usepackage{amsmath}
\usepackage{amssymb}
\usepackage{xcolor}
\usepackage[left]{lineno}
\usepackage{mhchem}

\usepackage{lipsum}
\def\MBT{MnBi$_2$Te$_4$}

\usepackage{comment}

\begin{document}

\title{Magneto-optical Kerr effect in an A-type antiferromagnet}

\author{V. Sunko}
\email{vsunko@berkeley.edu}

\affiliation {Department of Physics, University of California, Berkeley, California 94720, USA}
\affiliation {Materials Science Division, Lawrence Berkeley National Laboratory, Berkeley, California 94720, USA}
\affiliation {Institute of Science and Technology Austria, Am Campus 1, 3400 Klosterneuburg, Austria }

\author{S. Ahsanullah}
\author{V. Jain}
\affiliation {Department of Physics and Astronomy, University of Kansas, Lawrence, Kansas 66045, United States}

\author{S. Weber}
\affiliation {Department of Physics, Chalmers University of Technology, 412 96 Göteborg, Sweden}

\author{S. Kumaran}
\affiliation {Cavendish Laboratory, University of Cambridge, JJ Thomson Avenue, Cambridge CB3 0HE, United Kingdom}

\author{J.-Q. Yan}
\affiliation {Materials Science and Technology Division, Oak Ridge National Laboratory, Oak Ridge, Tennessee 37831, USA}

\author{J. Orenstein}
\email{jworenstein@lbl.gov}
\affiliation {Department of Physics, University of California, Berkeley, California 94720, USA}
\affiliation {Materials Science Division, Lawrence Berkeley National Laboratory, Berkeley, California 94720, USA}

\author{D. Ovchinnikov}
\email{d.ovchinnikov@ku.edu}
\affiliation {Department of Physics and Astronomy, University of Kansas, Lawrence, Kansas 66045, United States}

\begin{abstract}

\textbf{Magneto-optic Kerr effect (MOKE) is a powerful probe of broken time-reversal symmetry ($\mathcal{T}$), typically used to study ferromagnets. While MOKE has been observed in some antiferromagnets (AFMs) with vanishing magnetization, it is often associated with structures whose symmetry is lower than basic collinear, bipartite order. In contrast, theory predicts a mechanism for MOKE intrinsic to all AFMs of A-type, i.e. layered AFMs in which ferromagnetic layers are antiferromagnetically aligned. Here we report the first experimental confirmation of this mechanism in a bulk AFM. We achieve this by measuring the imaginary component of MOKE as a function of photon energy in MnBi$_2$Te$_4$, an A-type AFM where $\mathcal{T}$ is preserved in combination with a translation. By comparing the experimental results with model calculations, we demonstrate that observable MOKE should be expected in all collinear A-type AFMs with out-of-plane spin order, thus enabling optical detection of AFM domains and expanding the scope of MOKE to few-layer AFMs.}
\end{abstract}

\maketitle

The detection of broken time-reversal symmetry ($\mathcal{T}$) in quantum materials is of fundamental interest and practical relevance. Recent advances in the synthesis and exfoliation of van der Waals magnets have highlighted the critical role of optical techniques in detecting $\mathcal{T}$-breaking in samples just a few atomic layers thick~\cite{Gong_Discovery_2017, Huang_layer_2017,Mak_probing_2019, fei_two-dimensional_2018-1, gibertini_magnetic_2019, huang_emergent_2020,   zhao_emergence_2024, sun_resolving_2025, park_ferromagnetism_2025}, whose small mass poses a challenge for bulk probes of magnetism. The magneto-optic Kerr effect (MOKE), which is the difference in reflectivity for left and right circularly polarized  (LCP and RCP) light, is routinely used for detection of ferromagnetic (FM) order. Which classes of antiferromagnetic (AFM) order yield a measurable MOKE signal, and the underlying mechanism in each case, are questions of long-standing fascination. Here we show that MOKE is a more powerful and general probe of AFM order than previously thought, with implications for research and applications of both bulk and few-layer AFMs.  

\begin{figure}
    \centering
    \includegraphics[width=0.55\linewidth]{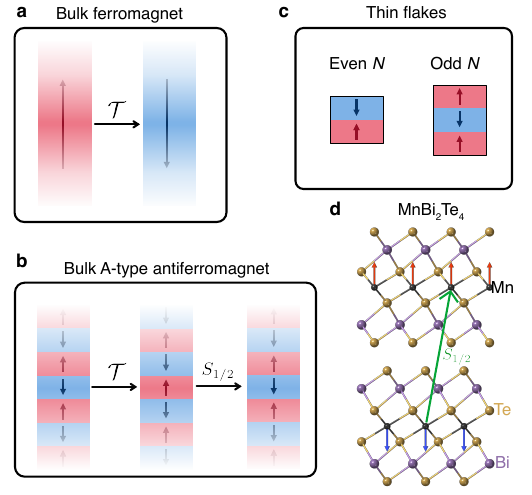}
    \caption{\textbf{Schematics of representative magnetic structures}. (a) Ferromagnet, where $\mathcal{T}$ reverses the sign of the order parameter; (b) A-type antiferromagnet, where the sign reversal  by $\mathcal{T}$ can be undone by a translation by half a magnetic unit cell ($S_{1/2}$); (c) Thin flakes of an A-type antiferromagnet: flakes of odd layer number $N$ have a net magnetic moment, while those with even $N$ do not. (d) Two layers of the \MBT~ crystal and magnetic structure, demonstrating the A-type AFM phase. Opposite spins are related by $S_{1/2}$.}
    \label{fig:Fig1}
\end{figure}

Time reversal flips the direction of spin, turning a ferromagnet into its time-reversed counterpart (Fig.~\ref{fig:Fig1}a). The two configurations with opposite magnetization $M$, called domains, enable the storage, processing, and retrieval of information in the form of a classical 0 or 1. Antiferromagnetic (AFM) order is also characterized by degenerate ground states related by $\mathcal{T}$, but $M=0$ in each domain. In the simplest example neighboring spins are antiparallel (Fig.~\ref{fig:Fig1}b). AFMs play an integral role in critical technologies~\cite{ baltz_antiferromagnetic_2018, nemec_antiferromagnetic_2018}, and interest in their properties has continued to grow~\cite{han_coherent_2023, chen_emerging_2024, smejkal_anomalous_2022, weber_surface_2024}. Zero magnetization can be a feature, as the absence of the long-range magnetic dipolar interaction leads to enhanced scalability.

A non-zero $M$ induces a difference in the index of refraction of LCP and RCP light, yielding a mechanism for MOKE that is essentially the same in all FMs. MOKE is therefore a reliable technique for distinguishing FM domains with high-spatial resolution.   In contrast, domains have been detected through MOKE only in a few AFMs~\cite{krichevtsov_spontaneous_1993, higo_large_2018, wu_magneto-optical_2020}, and the mechanism that underlies this signal has been attributed to specific properties of individual compounds.

Here we focus on a class of AFMs in which the time-reversed states are related by translation by half the magnetic unit cell, $S_{1/2}$, illustrated in bulk and thin flake forms in Figs.~\ref{fig:Fig1}b and c, respectively. We choose this class because the product $\mathcal{T} S_{1/2}$ ensures that the index of refraction of LCP and RCP light is the same as they propagate through an unbounded medium, a condition that at first sight seems to stipulate the absence of MOKE, since it is the difference of the two indices that causes MOKE in FMs. However, this conclusion is incorrect: MOKE by its nature involves a bounding surface that breaks $\mathcal{T} S_{1/2}$, unless $S_{1/2}$ is parallel to the surface. Therefore, $\mathcal{T} S_{1/2}$ imposes no constraints on MOKE at surfaces that break translational invariance described by $S_{1/2}$.

We investigate \MBT, a layered topological AFM with a N\'{e}el temperature of \unit[25]{K}~\cite{otrokov_prediction_2019}, as an example of a $\mathcal{T} S_{1/2}$ - symmetric AFM. Spins in individual Mn layers in \MBT~ are parallel to each other, and antiparallel to the spins in neighboring layers~(Fig.~\ref{fig:Fig1}d). This order, called A-type, persists to the surface of~\MBT~\cite{sass_robust_2020}. MOKE was recently reported in the wavelength range of $\unit[500-1000]{nm}$ in thin flakes of \MBT,  in samples with an odd number of layers ($N$), which have a net magnetization, and in samples with an even $N$, which do not~\cite{ovchinnikov_intertwined_2021, yang_odd-even_2021, qiu_axion_2023, bartram_real-time_2023}. While symmetry permits MOKE for all $N$, it does not identify the underlying mechanism. 

Axion electrodynamics associated with topology was proposed as the mechanism for MOKE in~\MBT~\cite{qiu_axion_2023, ahn_theory_2022-1}. The discontinuity in the topological $\mathcal{Z}_2$ invariant at the sample/vacuum interface manifests as a quantized surface Hall conductance, which is indistinguishable from a quantized trace of the magnetoelectric (ME) tensor in the static limit~\cite{essin_magnetoelectric_2009-1}. The surface conductance gives rise to a traceful, non-quantized ME tensor above zero frequency~\cite{lei_kerr_2023, ahn_theory_2022-1}. Optical phenomena associated with a traceful ME tensor  (collectively known as axion electrodynamics~\cite{Wilczek_two_1987}), provide a mechanism for MOKE~\cite{hornreich_theory_1968, orenstein_optical_2011, ahn_theory_2022-1}, as demonstrated in studies of the non-topological ME Cr$_2$O$_3$~\cite{ krichevtsov_spontaneous_1993}. 

A  distinct mechanism for MOKE in $\mathcal{T} S_{1/2}$ symmetric AFMs was proposed by Dzyaloshinskii and Papamichail~\cite{dzyaloshinskii_nonreciprocal_1995}. They treated the layered AFM as a stack of FMs with alternating spin direction. The dielectric tensor of each layer is assumed to be the same as in an unbounded medium with the same FM order, and MOKE arises due to the variation in dielectric properties as the wave propagates. In the `alternating FM' framework, the optical response arises from the bulk magnetic structure, and the surface conductance does not play a special role, unlike in the axion electrodynamics mechanism, where the surface contribution is essential. To the best of our knowledge, MOKE arising from the alternating FM effect has not been experimentally recognized prior to this work. 

Below we report measurements of MOKE in `infinite' ($\sim\unit[136]{nm}$ thick, Fig.~\ref{fig:SamplePhoto}) layer \MBT~as a function of magnetic field, $H$, and wavelength of the optical probe, $\lambda$. We find that non-zero MOKE appears in the $M = 0$ state, and that its spectrum is quantitatively captured by the alternating FM model~\cite{dzyaloshinskii_nonreciprocal_1995}, without invoking axion electrodynamics. The alternating FM mechanism is allowed in all collinear AFMs whose spins are oriented perpendicular to the layers, regardless of the number of layers, band topology or symmetry, opening the door for MOKE studies of a broader class of AFMs than was previously considered feasible.

\section*{The experiment: Magnetic field, position and wavelength dependence}

\begin{figure}[t]
    \centering  \includegraphics[width=0.525\linewidth]{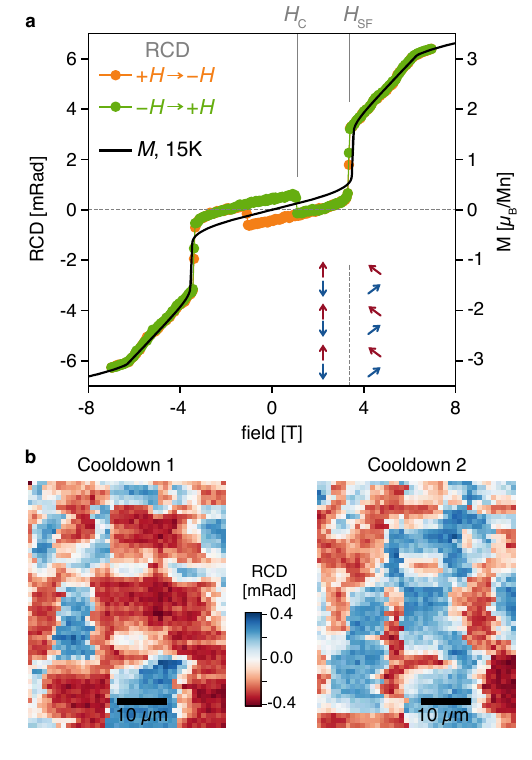}
\caption{\textbf{Magneto-optical measurements of MnBi$_2$Te$_4$.} (a) Reflection circular dichroism (RCD) measured at $\unit[525]{nm}$ as a function of magnetic field, showing a jump at the spin-flop transition ($H_{SF} \approx \unit[3.4]{T}$) and a hysteresis loop with a coercive field of $H_C \approx \unit[1]{T}$. A small setup-induced background contribution has been subtracted (Fig.~\ref{fig:BackgroundSubtraction}). RCD is compared with bulk magnetization measurements taken at $\unit[15]{K}$ with the field applied along the crystallographic $c$-axis (black line). The inset illustrates the bulk spin-flop transition. (b) Spatially resolved RCD measurements at $H=0$, revealing AFM domain structures that change between cooldowns.}
    \label{fig:Fig2}
\end{figure}

The complex MOKE angle is defined by the relation, $\Theta\equiv -i \delta r/r$, where the reflectivities of incident LCP and RCP light are given by $r=r_0\pm\delta r$ (the sign and polarization conventions are defined in the Methods section). Throughout this paper we show measurements of the imaginary part of $\Theta$, which is known as reflection circular dichroism (RCD). We measure RCD using a photoelastic modulator to vary the helicity of the incident light at $\unit[50]{kHz}$, using a setup illustrated in Fig.~\ref{fig:setup}. The change in reflectivity synchronous with the helicity modulation is proportional to RCD (see Methods for details of the experiment). 

In Fig.~\ref{fig:Fig2}a we compare the magnetic field dependence of RCD to that of bulk magnetization at $\unit[15]{K}$, measured on a sample from the same batch. RCD reveals two discontinuous features: a jump at $H_{SF}\approx\unit[3.4]{T}$ and a hysteresis loop with a coercive field of $H_{C}\approx\unit[1]{T}$. The jump corresponds to a spin-flop (SF) transition, illustrated in the inset, which is also seen in magnetization. The corresponding magnetization increase is $\Delta M_{SF}\approx 1.2 \left(\mu_B/\text{Mn}\right) \approx 0.4 M_s$, where $M_s = 3 \left(\mu_B/\text{Mn}\right)$ is the ordered moment at $\unit[15]{K}$~\cite{yan_crystal_2019}. The hysteresis loop has no counterpart in bulk magnetization measurements, which find $M=0$ for $H<H_{SF}$. Therefore, the two non-zero values of RCD at $H=0$ correspond to time-reversed states, each with $M=0$. Further evidence for this interpretation is provided by spatially resolved measurements (Fig.~\ref{fig:Fig2}b) that reveal spontaneous formation of domains with positive and negative RCD values in zero field, with distinct structures in different $H=0$ cooldowns. 

\begin{figure*}[t]
    \centering
    \includegraphics[width=0.9\linewidth]{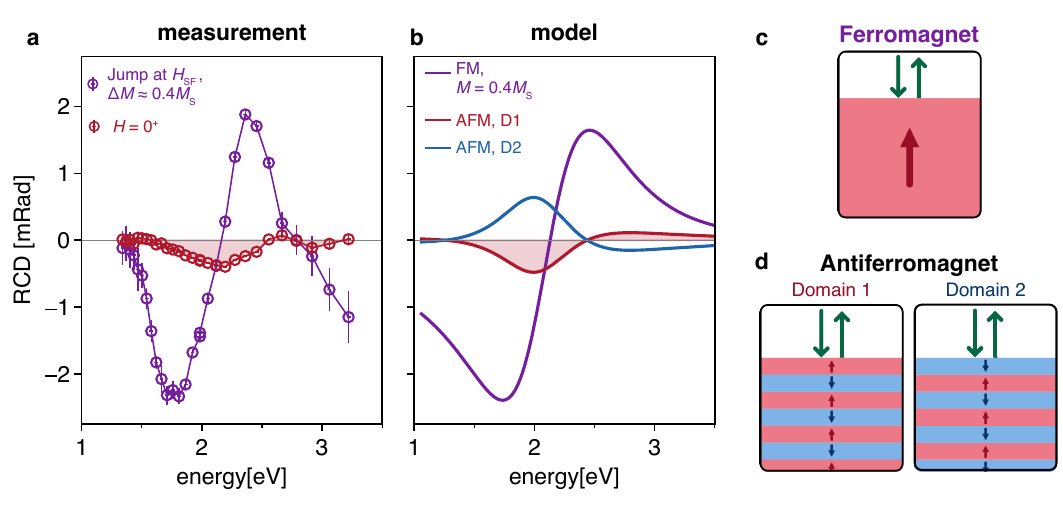}
    \caption{\textbf{Spectroscopic signature of MOKE in antiferromagnetic MnBi$_2$Te$_4$.} (a) Measured RCD spectrum at the spin-flop transition (purple) and at zero field (red), showing distinct spectral shapes. For each photon energy, we measured the magnetic field dependence of RCD (Fig.~\ref{fig:WavelengthFieldDependence}). The AFM spectrum is half of the difference between RCD on the downward ($H=0^{+}$) and upward ($H=0^{-}$) field sweep, averaged across the hysteresis loop, between $ \pm \unit[0.9]{T}$. The SF spectrum is given by $\Theta_{SF}=\Theta(\unit[3.4]{T})-\left[\Theta(\unit[3.2]{T})-\Theta(0^{+})\right]$. The error bars for the $H=0^{+}$ measurement represent the standard deviation across this field range. The larger error bars for the SF spectrum reflect the fact fewer measurements were used to extract the corresponding RCD value (at $H=\unit[3.4]{T}$ and $H=\unit[3.2]{T}$). Laser power fluctuations do not contribute to these error bars, since the RCD is normalized by the simultaneously measured reflectivity (the same detector voltage is demodulated at two frequencies). The only relevant systematic uncertainty is from the PEM calibration, which for our Hinds PEM-100 is $\leq 5\%$, corresponding to $<0.1$\,mrad on the RCD scale.} (b) RCD spectra computed from the Lorentz model (Eq.~\ref{eq:Lorentz}, with parameters from Table~\ref{tab:optical_params}) for a ferromagnet (purple, schematic in panel c) and two antiferromagnetic domains (red and blue, schematics in panel d). The ferromagnetic RCD is scaled by a factor of $0.4$ from that calculated for a saturated moment, to be directly comparable to the measured RCD jump at the spin-flop transition.

    \label{fig:Fig3}
\end{figure*}

The existence of two discontinuities in RCD \textit{vs.} $H$ provides an unusual opportunity to compare the MOKE spectrum in a phase with $M\neq0$, that is the spin-flop (SF) phase,  and a $\mathcal{T} S_{1/2}$ -symmetric AFM ($M=0$) phase in the same material. We measured the magnetic field dependence of RCD in the photon energy range from $\unit[1.34]{eV}$ ($\unit[925]{nm}$) to $\unit[3.22]{eV}$ ($\unit[385]{nm}$) (Fig.~\ref{fig:WavelengthFieldDependence}), and extracted RCD spectra in the two phases. In Fig.~\ref{fig:Fig3}a we compare the RCD spectrum corresponding to the SF transition with the RCD spectrum in the AFM phase. The AFM spectrum is half of the difference between RCD at $H=0^{+}$ and $H=0^{-}$,  where $H=0^{+}$ and $H=0^{-}$ denote zero field as approached from positive and negative fields. The SF spectrum, corresponding to the jump in magnetization across the transition, is given by $\Theta_{SF}=\Theta(\unit[3.4]{T})-\left[\Theta(\unit[3.2]{T})-\Theta(0^{+})\right]$, because $\Theta(\unit[3.2]{T})$ is a sum of the antiferromagnetic and magnetization contributions, while in the spin flop phase, at $\unit[3.4]{T}$, there is no antiferromagnetic contribution.

The two spectra are strikingly different, ruling out a small uncompensated bulk ferromagnetic moment as a trivial explanation for RCD at $H=0$. Further, the fact that the two spectra are rich in features creates an opportunity to constrain theories: a suitable theoretical model should simultaneously reproduce both of them.

\section*{Lorentz model: ``infinite" and ``finite" samples}

The RCD spectrum measured in the SF state is consistent with resonant MOKE in a FM. Both the magnitude and zero crossing can be reproduced by a Lorentz oscillator whose resonant photon energy differs for LC and RC polarized light. The corresponding dielectric function is:
\begin{equation}\label{eq:Lorentz}
\varepsilon_{\pm}(\omega) = \varepsilon_\infty + \frac{f}{(\omega_0 \pm\delta\omega)^2 - \omega^2 - i\gamma \omega},
\end{equation}
where $f$, $\omega_0$, and $\gamma$ are oscillator strength, resonant frequency, and damping. These parameters, together with a background contribution, $\varepsilon_\infty$, are chosen to reproduce the measured optical conductivity in the spectral range of interest, reported in Ref.~\cite{kopf_influence_2020}. The good agreement between the measured quantities and those calculated from the Lorentz model is shown in Fig.~\ref{fig:LorentzExpComparison}. Finally, $\delta\omega$, the difference in resonance frequency for LC and RC polarized light, is chosen to fit the magnitude of the measured RCD.  The purple curve in Fig.~\ref{fig:Fig3}b shows that the single oscillator model captures the main features of the observed ferromagnetic RCD spectrum (Fig.~\ref{fig:Fig3}a). 

Our quantitative description of RCD in the ferromagnetic phase of \MBT~leads us to consider the alternating FM model for the RCD in its AFM state: a stack of FM layers in which the sign of $\delta \omega$ alternates in neighboring layers. We use the transfer matrix formalism~\cite{zak_universal_1990}, following the approach in Ref.~\cite{hendriks_enhancing_2021}, to calculate the reflection coefficient from the stack. The number of layers is 100, and the thickness of each layer, $d=\unit{1.3}{nm}$, corresponds to the separation between spins in \MBT. The simulated RCD spectra for the two $\mathcal{T}$-reversed AFM domains (Fig.~\ref{fig:Fig3}d), are shown in red and blue in Fig.~\ref{fig:Fig3}b. Remarkably, this minimal model captures the spectral shape and the magnitude of RCD in the AFM phase of \MBT, without introducing free parameters beyond those that describe the FM phase. In the SI we extend this analysis by using a dielectric function computed from a density functional theory (DFT) calculation to construct the alternating FM model, and we find that this approach also yields the correct magnitude of RCD in the AFM phase (Fig.~\ref{fig:DFTModels_offset}). To the best of our knowledge, this is the first experimental evidence for an RCD arising from an alternating FM model in a bulk AFM.

\begin{figure*}[t]
    \centering
    \includegraphics[width=0.9\linewidth]{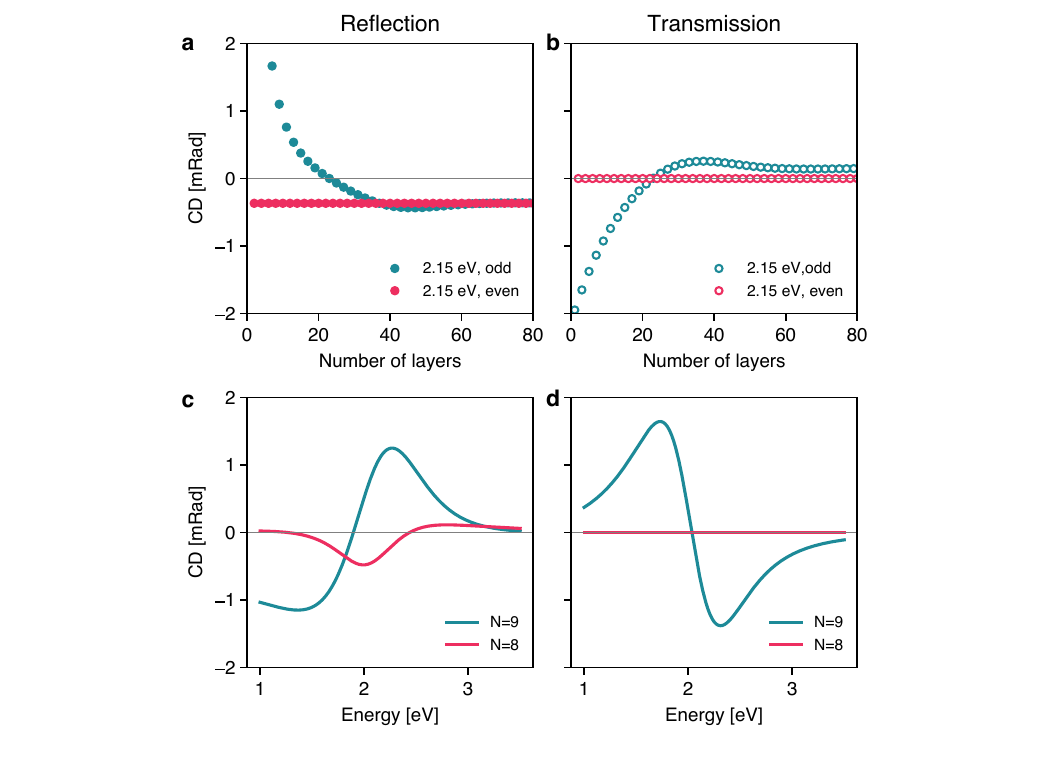}
    \caption{\textbf{Dependence of circular dichroism on the number of layers} (a, b) Calculated reflection (RCD) and transmission (TCD) circular dichroism as a function of layer number for even and odd layer stacks, using the same model and parameters as in Fig.~\ref{fig:Fig3}. (c, d) RCD and TCD spectra for 8-layer (8L) and 9-layer (9L) AFM stacks. }
    \label{fig:Fig4}
\end{figure*}

Next, we show that the key features of RCD and transmission circular dichroism (TCD) that are seen in few layer crystals in Ref.~\cite{qiu_axion_2023} can be understood within the same model. We performed a series of revealing transfer matrix calculations, varying the number of layers while keeping the thickness of individual layers fixed. In Figs.~\ref{fig:Fig4}(a,b) we plot the calculated RCD and TCD, respectively, as a function of layer number at fixed photon energy. We find that both RCD and TCD depend strongly on $N$ if it is odd, but are independent of $N$ when it is even, consistent with the findings of Ref.~\cite{dzyaloshinskii_nonreciprocal_1995}. When the slab thickness $Nd$ far exceeds the optical penetration depth, RCD in even and odd layer samples converge to the same value, as expected. TCD vanishes for all even layer stacks, a phenomenon that can be understood from a symmetry perspective: even layer stacks are symmetric under a product of inversion and  time-reversal symmetry, which prohibits circular dichroism in transmission~\cite{dzyaloshinskii_space_1991,canright_ellipsometry_1992, armitage_constraints_2014, qiu_axion_2023}.

Further, the dichroism spectra for even and odd layer stacks are strikingly different (Fig.~\ref{fig:Fig4}c, d). For $N=8$, the RCD spectrum is identical to the spectrum of the bulk AFM crystal (Fig.~\ref{fig:Fig3}b), while spectrum for $N=9$ resembles that of the bulk SF phase, as $M\neq0$ for odd layer stacks. The maximum RCD is about three times larger for $N=9$ than for $N=8$. However, these calculations assume that samples are suspended in vacuum, which is not a realistic experimental scenario. The transfer matrix approach allows us to include the substrate in the calculation of reflectivity~\cite{hendriks_enhancing_2021}, and we find that doing so preferentially suppresses dichroism in odd-layer samples. In Fig.~\ref{fig:Sapphire} we show that the maximum magnitude of RCD for samples on sapphire and diamond substrates, used in Ref.~\cite{qiu_axion_2023}, is approximately equal in magnitude for even and odd layer stacks, while they maintain the characteristic spectral shapes shown in Fig.~\ref{fig:Fig4}c. We therefore find that the alternating FM model with realistic parameters captures the main experimental results on $N$-even and $N$-odd flakes of \MBT~reported in~\cite{qiu_axion_2023}, as well as on `infinite' layer flakes reported here.

\section*{Bilayer: the exact solution and the physical picture}

While the numerical results shown in the above section capture the experimental results, the underlying physics is somewhat mysterious, and poses the following questions: why does MOKE arise in $\mathcal{T} S_{1/2}$-symmetric AFMs within the alternating FM model, and what influences its magnitude? Why is its value independent of the (even) number of layers? Why does TCD vanish? To address these questions, we first consider the analytical solution for MOKE in a bilayer, and then offer an intuitive physical picture, shedding light on the experimental, numerical and analytical findings. 

\begin{figure*}[t]
    \centering
    \includegraphics[width=1\linewidth]{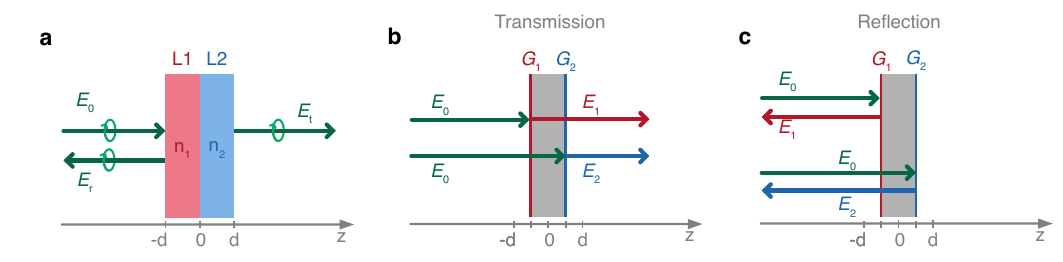}
    \caption{\textbf{Illustration of the model for MOKE in an antiferromagnetic bilayer.} (a) Schematic representation of a bilayer, consisting of two layers of indices of refraction $n_{1,2}=n\pm\delta n$ when probed with circularly polarized light. $\delta n$ is induced by the opposite magnetization in the two layers, $\pm M_z$. (b,c) Simplified representation of a bilayer, where the layers are replaced by two infinitesimally thin sheets of conductances $G_{1,2}=G_0\pm\delta G$, separated by a dielectric medium of thickness $d$ and index of refraction $n$. The  propagation-induced differences between fields emitted from the two sheets cancel in (b) transmission, but not in (c) reflection, leading to vanishing TCD and non-zero RCD.}
    \label{fig:Fig5}
\end{figure*}

We consider a single bilayer of oppositely oriented FMs. Each layer is characterized by the index of refraction $n+\delta n$, where the sign of $\delta n$ is opposite for the two layers (Fig.~\ref{fig:Fig5}a). First, we utilize the transfer matrix approach to reproduce the expressions for the transmission and reflection coefficients of such a bilayer, derived in Ref.~\cite{dzyaloshinskii_nonreciprocal_1995}.  MOKE in the physically relevant limit ($k_0d\ll1$ and $\delta n\ll n$) is given by

\begin{equation}\label{eq:Kerr}
    \Theta_{AFM} = -k_0nd\frac{2 \delta n} {n^2-1}= \left(-ik_0nd\right) \Theta_{FM},
\end{equation}

\noindent where $\Theta_{FM}$ is the Kerr angle of the corresponding bulk ferromagnet, $k_0=\omega/c$ the wavevector in vacuum, and $n$ the average index of refraction of the two layers. The expression for arbitrary layer thickness and $\delta n$ is given in the Supplementary Information.

Eq.~\ref{eq:Kerr} showcases the common microscopic origin of the MOKE response of an AFM bilayer and that of the corresponding ferromagnet, but it also reveals a surprising aspect of the link between them.  It is natural to assume that a $\mathcal{T} S
_{1/2}$ - symmetric type-A AFM can exhibit nonzero MOKE only if strong absorption (governed by the imaginary part of $n$) leads to disproportionate sampling of the top layer. However, Eq.~\ref{eq:Kerr} shows that this is not correct: the magnitude of $\Theta_{AFM}$ depends on the absolute value of the index, $|n|$, and is nonzero even in the absence of dissipation. The phase of $n$ appears in the phase shift between $\Theta_{AFM}$ and $\Theta_{FM}$, which manifests as the difference in the spectrum of RCD between AFM and FM structures, as seen for example in Fig.~\ref{fig:Fig3}a.

To gain further insight into the physical mechanism behind the alternating FM scenario, we analyze a toy model (Fig.~\ref{fig:Fig5}(b,c)) in which each of the two layers is replaced by an infinitesimally thin ferromagnetic sheet of optical conductance $G_{1,2}$, where $1,2$ are layer indices. The sheets are separated by a non-magnetic medium of thickness $d$ and index of refraction $n$. For illustrative purposes we analyze a single reflection from each sheet. 

We consider a circularly polarized plane wave propagating in the positive $z$ direction with amplitude $E_0$. The electric field generates sheet currents,
\begin{equation}\label{eq:K12}
    K_{1,2}=G_{1,2} E_0  e^{\mp ik_0 n d/2},
\end{equation}
where we have taken $z=0$ as the midpoint between the sheets for convenience. The currents $K_{1,2}$ radiate, emitting electric fields, 

\begin{equation}\label{eq:E12}
    E_{1,2}(z) = \frac{\mu_0c}{2} K_{1,2}  e^{ik_0 \left|z\pm d/2\right|},
\end{equation}
where $c$ is the speed of light and $\mu_0$ the vacuum permeability. 

The total transmitted electric field, given by the sum of $E_{1}(z)$ and $E_{2}(z)$ for $z>d/2$, is, 

\begin{equation}\label{eq:Tfield}
E_T(z)=\mu_0cE_0 \left(\frac{G_1 +G_2}{2}\right) e^{ik_0z}=\mu_0cE_0G_{0}e^{ik_0z}.
\end{equation}
As $E_T(z)$ depends only on the average conductance $G_0$, the TCD is zero. The toy model shows that this is a consequence of cancellation of propagation factors. The incident wave propagates an additional distance $d$ in the medium beyond layer 1 to induce currents layer 2. However, the wave emitted by layer 1 must propagate through the medium to reach a given point $z>d/2$. The total propagation factor, $e^{iknz}$, for waves transmitted through the bilayer is the same, and independent of $d$ (Fig.~\ref{fig:Fig5}b). The sum of  fields emitted from the two sheets therefore depends only on the sum of the currents, $G_{1}+G_{2}$. 

In contrast, the propagation factors for fields radiated in the $-z$ direction add, rather than cancel: the incident EM wave travels further to excite current in layer 2 and its radiated field propagates further before reaching an observer at $z<-d/2$ (Fig.~\ref{fig:Fig5}c). The total reflected electric field is given by, 

\begin{equation}\label{eq:Rfield}
    \begin{aligned}
        E_{R}(z) = & \frac{\mu_0cE_0}{2} e^{-ik_0z} \left(G_{1} e^{-ik_0nd} +G_{2} e^{ik_0nd}\right) \\
        & \approx  \mu_0cE_0 e^{-ik_0z} \left(G_{0}-\delta G ik_0nd\right),
    \end{aligned}
\end{equation}
where we used $|k_0nd|\ll 1$ in the expansion of the exponential. The term proportional to $\delta G$ gives rise to $\Theta_{AFM}$. 

The toy model for a bilayer identifies the origin of nonzero $\Theta_{AFM}$ as the change in the electric field as it propagates, regardless of whether it is dominated by the real or imaginary component of $n$. We see that absorption does not play a special role in generating MOKE. Further, RCD does not depend on the layer number for even layers (Fig.~\ref{fig:Fig4}a) because of the exponential nature of propagation: the ratio of fields in two neighboring layers is $\exp{(ik_0nd)}$, independent of the number of layers. Thus the effects that yield RCD in two atomic layers and in a semi-infinite AFM crystal are the same. These qualitative features found in the toy model are retained when multiple reflections are included through the transfer matrix approach (Eq.~\ref{eq:Kerr}, and Fig.~\ref{fig:Fig4}).

\section*{Conclusion}

In this paper we identify alternating FM layers as the origin of MOKE in an A-type antiferromagnet with $\mathcal{T} S_{1/2}$ symmetry, therefore providing first experimental confirmation of the mechanism proposed by Dzyaloshinskii and Papamichail~\cite{dzyaloshinskii_nonreciprocal_1995}. Within this theory, $\Theta_{AFM}$ is entirely determined by the MOKE response of the ferromagnetic layers that are the building blocks of the AFM, allowing us to test it directly by tuning \MBT~ across the spin-flop transition,  thus accessing $\Theta_{AFM}$ and $\Theta_{FM}$ in the same crystal. Based on the quantitative agreement between the measured RCD spectrum in the AFM phase and the theoretical prediction based solely on the FM spectrum, we conclude that MOKE observed at visible and near infrared wavelengths in \MBT~arises from alternating FM layers, rather than axion electrodynamics associated with band topology. Axion electrodynamics may dominate the MOKE response at terahertz frequencies below the bulk band gap~\cite{lei_kerr_2023, ahn_theory_2022-1, han_quantized_2025, wu_quantized_2016}, where there are no other optical transitions that could overwhelm their contribution, although results in even-layer flakes suggest that the axion contribution in this frequency range is weak~\cite{lei_kerr_2023, han_quantized_2025}.

The effect we observe is not limited to \MBT, and is generically present in A-type antiferromagnets with spins out of plane. We believe that it has not been recognized experimentally in part because the signal vanishes when averaged over atomic steps, naturally present in as-grown bulk crystals. However, optical experiments in van der Waals materials regularly probe surfaces that are atomically flat across the length scale of a laser beam. We note that some previous sightings of MOKE in van der Waals AFMs with an even number of layers were conditionally attributed to extrinsic magnetization~\cite{Huang_controlling_2018, zhang_gate-tunable_2020,yang_odd-even_2021}, given the widespread conviction that MOKE intrinsic to A-type AFMs was forbidden by symmetry. While this explanation is certainly plausible, our findings show that intrinsic MOKE in A-type AFMs is always allowed, and other mechanisms should be considered when $\Theta_{AFM}$ cannot account for the magnitude, or the spectrum, of the observed signal. 

Correctly estimating $\Theta_{AFM}$ is therefore essential for interpretation of experimental results. Our work suggests a strategy to do so, both in bulk and thin flakes. In bulk, it is sufficient to combine measured $\Theta_{FM}$ and optical conductivity with the analytical expression for $\Theta_{AFM}$ (Eq.~\ref{eq:Kerr}). For thin flakes, it is important to consider the role of the substrate, which may suppress or enhance observed signals~\cite{hendriks_enhancing_2021}; we suggest the transfer matrix approach as a convenient way to do so. We note that $|\Theta_{AFM}|$ is not generically small compared to the noise floor of modern experimental setups: it is suppressed with respect to $|\Theta_{FM}|$ only by a factor of $|k_0nd|=|2\pi n d/\lambda|$, which is more than an order of magnitude larger than $d/\lambda$ suggested by dimensional analysis. 

Our findings show that MOKE is a powerful and general tool for investigating $\mathcal{T} S_{1/2}$ symmetric AFMs. In this work we used it to image time-reversed domains, as shown for in Fig.~\ref{fig:Fig2}b.

Further, the nonzero $\Theta_{AFM}$ enabled the discovery that a magnetic field can select antiferromagnetic domains (Fig.~\ref{fig:Fig2}a) in such magnets, challenging the conventional view that antiferromagnets cannot be controlled by external magnetic fields. Uncovering the mechanism behind field control is beyond the scope of this study, but we note that the weakened exchange interaction at the surface is a plausible cause~\cite{sass_robust_2020, chong_intrinsic_2024-1}. This implies that details of the surface can affect the coercive field, as confirmed by measurements on two flakes exfoliated from the same batch (Fig.~\ref{fig:Two_flakes}). However, the magnitude of the RCD at $H = 0$ is a bulk property, independent of local surface conditions. The coercivity and field-switching process will be the subject of future computational and experimental investigations, enabled by the discovery of the AFM RCD signal. In summary, our results address fundamental questions concerning the origin of MOKE in AFMs, provide a strategy for studying and controlling their domains, and elucidate results on exfoliated van der Waals AFMs.

\begin{acknowledgments}

We thank Christine Kuntscher for providing optical conductivity and reflectance data published in Ref.~\cite{kopf_influence_2020}, and Nicola Spaldin, Joel Moore and Bevin Huang for useful discussions.  

V.S. and J.O.  received support from the Gordon and Betty Moore Foundation's EPiQS Initiative through Grant GBMF4537 to J.O. at UC Berkeley. Experimental and theoretical work at LBNL and UC Berkeley was funded by the Quantum Materials (KC2202) program under the U.S. Department of Energy, Office of Science, Office of Basic Energy Sciences, Materials Sciences and Engineering Division under Contract No. DE-AC02-05CH11231. Work at the University of Kansas was supported by the U.S. Department of Energy, Office of Science, Basic Energy Sciences, EPSCoR, and Materials Sciences and Engineering Division under Award No. DE-SC0025319. Parts of device fabrication were performed in the KU Nanofabrication Facility, which is supported by the National Institutes of Health NIGMS P30GM145499. Work at ORNL was supported by the U. S. Department of Energy, Office of Science, Basic Energy Sciences, Materials Sciences and Engineering Division. For the DFT calculations we used resources provided by the Swedish National Infrastructure for Computing (SNIC) at  C3SE. We acknowledge support from the US National Science Foundation (NSF) Grant Number 2201516 under the Accelnet program of Office of International Science and Engineering (OISE). This publication is funded in part by a QuantEmX grant from ICAM and the Gordon and Betty Moore Foundation through Grant GBMF9616 to Sivaloganathan Kumaran.

\end{acknowledgments}

\bibstyle{apsrev4-1}
\bibliography{MBT_References}

\appendix

\newpage
\newpage

\counterwithout{equation}{section}
\renewcommand\theequation{M\arabic{equation}}
\renewcommand\thefigure{M\arabic{figure}}
\renewcommand\thetable{M\arabic{table}}
\renewcommand\thesection{M\arabic{section}}
\renewcommand\bibnumfmt[1]{[M#1]}
\setcounter{equation}{0}
\setcounter{figure}{0}
\setcounter{enumiv}{0}

\section*{Methods}
\subsection*{Complex MOKE: definitions}

As discussed in the main text, MOKE is a manifestation of difference in reflectivity between the light of two circular polarizations. The Jones vectors corresponding to the circular polarizations are given by: 
\begin{align}
    \mathbf{\hat{e}_+} &= \frac{1}{\sqrt{2}}\begin{pmatrix} 1 \\ i \end{pmatrix}, \quad \text{(LCP for \( k_z > 0 \), RCP for \( k_z < 0 \))} \label{eq:circularL} \\
    \mathbf{\hat{e}_-} &= \frac{1}{\sqrt{2}}\begin{pmatrix} 1 \\ -i \end{pmatrix}. \quad \text{(RCP for \( k_z > 0 \), LCP for \( k_z < 0 \))}, \label{eq:circularR}
\end{align}
capturing the fact that the LCP light becomes RCP upon reflection, and vice-versa. It is  convenient to use the $\mathbf{\hat{e}_\pm}$ basis, rather than the LCP and RCP basis, so that reflection from an ideal isotropic mirror is captured by the identity matrix. 

Vectors and matrices can be transformed between the  $\mathbf{\hat{e}}_{\pm}$ basis and basis of linear ($\mathbf{\hat{e}}_{x,y}$) polarization using the following transformation matrices:
\begin{equation}
    \mathbf{A}_{C \rightarrow L}= \frac{1}{\sqrt{2}}\begin{pmatrix} 1 & 1 \\ i & -i \end{pmatrix}, \quad \mathbf{A}_{L \rightarrow C}= \frac{1}{\sqrt{2}}\begin{pmatrix} 1 & -i \\ 1 & i \end{pmatrix}.
\end{equation} 

In the $\mathbf{\hat{e}_\pm}$ basis, the reflection matrix corresponding to circular dichroism is given by:
\begin{equation}\label{eq:reflectivityCircular}
    r_C =
    \begin{pmatrix}
        r_0+\delta r & 0\\
        0 & r_0 - \delta r
    \end{pmatrix},
\end{equation}

\noindent In the $\mathbf{\hat{e}}_{x,y}$ basis the reflectivity matrix is then equal to: 

\begin{equation}\label{eq:Rlinear}
    r_L =
    \begin{pmatrix}
        r_0 & r_{xy}\\
        -r_{xy} & r_0
    \end{pmatrix}= \begin{pmatrix}
        r_0 & -i\delta r\\
        i\delta r & r_0
    \end{pmatrix}.
\end{equation}
The complex Kerr angle is defined as:
\begin{equation}\label{eq:KerrDef}
    \Theta=\frac{r_{xy}}{r_0}=-i\frac{\delta r}{r_0},
\end{equation}
and RCD is the imaginary part of $\Theta$.

\subsection*{RCD Measurements}

The experimental setup is shown in Fig.~\ref{fig:setup}. The cryostat is Quantum Design Opticool, with a 7T magnet. Most measurements were done with a pulsed laser (output of Light Conversion Orpheus optical parametric amplifier, seeded by a Carbide laser, repetition rate $\unit[300]{kHz}$, pulse duration $\sim \unit[250]{fs}$), focused to a  spot of lateral dimensions of $\sim\unit[4]{\mu m}$. The position dependent measurements (Fig.~\ref{fig:Fig2}b) were done with a continuous wave laser, with wavelength $\unit[532]{nm}$, corresponding to the photon energy of $\unit[2.33]{eV}$, which we found to offer good signal to noise throughout the studied field range (Fig.~\ref{fig:Fig3}a). Spatial resolution is achieved by moving the sample underneath the focused laser spot using Attocube positioners. Regardless of the laser source, the average power of $\unit[1]{\mu W}$ reached the sample, the incident light was vertically polarized, and chopped using a mechanical chopper. 

The incident vertical polarization is rotated by $45\deg$ using a half wave plate. The next element in the optical path is the photoelastic modulator (PEM) set to a $1/4$ wave modulation, therefore modulating light polarization between linear and circular at $\unit[50]{kHz}$. The light intensity was measured using a Si photo-diode, through the current input of the Zurich Instruments lock-in amplifier. The intensity was simultaneously demodulated at the chopper and at the PEM frequencies, yielding measured intensities $I_c$ and $I_P$.  $\operatorname{Im}\left(\Theta\right)$ is found through their ratio as (see SI for a derivation):

\begin{equation}\label{eq:ImTheta}
    \operatorname{Im}\left(\Theta\right)= \frac{I_P}{I_c}\frac{1}{\pi J_1\left(\pi/2\right)}\approx\frac{I_P}{I_c}\frac{1}{1.78073},
\end{equation}
where $J_1$ is the Bessel function of first order.
Extracting $\operatorname{Im}\left(\Theta\right)$ through Eq.~\ref{eq:ImTheta} has the attractive feature that wavelength-dependent reflection and transmission properties of optical elements, as well as the responsivity of the photodiode, do not influence the measured spectrum, since they cancel out in the ratio $\frac{I_P}{I_c}$. Nonetheless, we took into account the wavelength dependence of the photodiode responsivity  and the transmission though the beamsplitter to ensure that the power incident on the sample is $\approx 1 \mu W$ across the spectral range. 

Imperfections of optical elements, coupled with the Faraday rotation the cryostat windows and objective, result in a spurious field- and wavelength- dependent additive background in the RCD measurements. We measured the background on a Si/SiO2 substrate, and subtracted it from measurements on \MBT~(Fig.~\ref{fig:BackgroundSubtraction}). Since the spectra in Fig.~\ref{fig:Fig3}a were obtained through differential measurements, it was not necessary to measure the background contribution for each photon energy.

The error bars in Fig.~\ref{fig:Fig3}a represent the statistical scatter of the RCD signal within the field range used to extract each value. For the AFM spectrum, the value at $H = 0$ is obtained by averaging the difference between down and up sweeps over the range $\pm 0.9$\,T, and the error bar is the standard deviation of the data points in that range. For the spin-flop spectrum, the value is taken from the difference between two field points (3.4\,T and 3.2\,T), so the standard deviation is correspondingly larger. Laser power fluctuations do not contribute, since the RCD is normalized by the simultaneously measured reflectivity (measurements were truly simultaneous, in that the same detector voltage is demodulated at two frequencies). The only relevant systematic uncertainty is from the PEM calibration, which here refers to the precision with which the modulation amplitude is known. We use a Hinds PEM-100, whose factory calibration guarantees retardation accuracy within 5\% for a narrow monochromatic beam at normal incidence, which is sufficient for our purposes, as this level of uncertainty corresponds to $<0.1$\,mrad on the RCD scale in Fig.~3a.

Since our measurements are done with the sample in vacuum, and there are no electrical contacts on the sample, all of the thermalization is happening through the Si/\ce{SiO2} substrate. This is very inefficient, leading to the known problem of large temperature gradients between the cryostat thermometer and the sample. Independently establishing the sample temperature is therefore difficult, but we use the bulk magnetization measurement as a `calibration' - the agreement between the RCD measurement and the magnetization proves that the sample is at ~18K.

\subsection*{Transfer matrix calculations}

The reflection and transmission coefficients for the AFM stack are found numerically through transfer matrix calculations, following the method detailed in Ref.~\cite{hendriks_enhancing_2021}. This method offers a computationally convenient way to satisfy boundary conditions on the electric and magnetic fields at each interface, and captures the propagation through a uniform material by a phase factor of $\exp{(i n k_0 d)}$, where $d$ is the thickness of the material, $n$ the complex index of refraction, and $k_0 = \omega/c$ is the wave vector in vacuum.

We formulate the numerical transfer matrix calculation in the basis of linear polarization. If the dielectric tensor is diagonal in the basis of circular polarization,

\begin{equation}
    \varepsilon_C =
    \begin{pmatrix}
        \varepsilon_+ & 0 \\
        0 & \varepsilon_-,
    \end{pmatrix}
\end{equation}
in the basis of linear polarization it is given by: 

\begin{equation}
    \varepsilon_L =
    \frac{1}{2}\begin{pmatrix}
        \varepsilon_+ + \varepsilon- & i(\varepsilon_- - \varepsilon+) \\
        -i(\varepsilon_- - \varepsilon+) & \varepsilon_+ + \varepsilon-
    \end{pmatrix}.
\end{equation}
This is the form of $\varepsilon$ we use for transfer matrix calculations, with $\varepsilon_{\pm}$ given by Eq.~\ref{eq:Lorentz}. 
The calculation will return the reflection matrix of the form given by Eq.~\ref{eq:Rlinear}, allowing us to calculate the complex Kerr angle through Eq.~\ref{eq:KerrDef}, with its imaginary part corresponding to the reflection circular dichroism. A matrix of similar form is found for transmission. 

We can use the same formalism to calculate the reflection and transmission from more complex structures. For instance, a calculation of reflection from a small $N$ sample suspended in vacuum is unrealistic since samples are necessarily placed on substrates. In Fig.~\ref{fig:Sapphire} we compare the reflection of a $N=8$ and $N=9$ layer sample suspended in vacuum to the same samples on sapphire and diamond substrates, as was done in experiments in Ref.~\cite{qiu_axion_2023}. We find that the $N=9$ reflection is suppressed more than the $N=8$ reflection, making the magnitude of the even and odd layer RCD very similar to each other.

\subsection*{Density functional theory}

For our density functional calculations of the frequency-dependent dielectric function for MnBi$_2$Te$_4$, we employ the Vienna \textit{ab-initio} software package (VASP)~\cite{kresse_efficient_1996}. We use the generalized gradient approximation (GGA) with the Perdew-Burke-Ernzerhof (PBE) functional~\cite{perdew_generalized_1996}. The Projector-augmented wave (PAW) method~\cite{blochl_projector_1994} with the standard VASP pseudopotentials is used, and we take Mn: 3d$^6$4s$^1$; Bi: 5d$^{10}$6s$^2$6p$^3$ and Te: 5s$^2$5p$^4$ electrons as valence. To approximately account for the localized nature of the unpaired d electrons in Mn, we use the DFT+U method~\cite{anisimov_first-principles_1997} and adopt the rotationally invariant method by Dudarev et al.~\cite{dudarev_electron-energy-loss_1998}. We set U=4 eV on the Mn atoms.
 
Given that MnBi$_2$Te$_4$ is a layered material, we implement van der Waals corrections, using the DFT-D3 method of Grimme et al.~\cite{grimme_consistent_2010}. We relax the bulk, primitive rhombohedral cell of MnBi$_2$Te$_4$ using a Gamma-centered mesh of 13$\times$13$\times$5 to sample the Brillouin zone and a kinetic energy cutoff of 600 eV for our plane-wave basis set. For the structural relaxation, we neglect spin-orbit coupling (SOC), and enforce ferromagnetic order of the Mn ions using spin-polarized collinear calculations. We use a Gaussian smearing for the partial occupancies with a smearing width of 0.01 eV.

To calculate the frequency-dependent dielectric tensor within the independent particle approximation (that is, neglecting local field effects), we use the method developed by Gajdoš et al.~\cite{gajdos_linear_2006}. For these calculations we include SOC self-consistently and enforce ferromagnetic ordering with the spin axis perpendicular to the Mn layers. Having first obtained the ground-state electronic density and corresponding Kohn-Sham states in a self-consistent DFT calculation, the imaginary part of $\varepsilon$ is calculated with these wavefunctions using the following equation:

\begin{align}
    \varepsilon^{(2)}_{\alpha\beta}(\omega) &= \frac{4\pi^2 e^2}{\Omega} \lim_{q \to 0} \frac{1}{q^2} \sum_{c,v,\mathbf{k}} 2w_{\mathbf{k}} \delta(\epsilon_{c\mathbf{k}} - \epsilon_{v\mathbf{k}} - \omega) \nonumber \\
    &\quad \times \langle u_{c\mathbf{k}+\mathbf{e}_\alpha q} | u_{v\mathbf{k}} \rangle \langle u_{c\mathbf{k}+\mathbf{e}_\beta q} | u_{v\mathbf{k}} \rangle^*,
\end{align}

Here,  $w_k$ is the weight of point $k$ in the discrete summation over the Brillouin zone, $\Omega$ is the volume of the unit cell, and the indices $c$ and $v$ denote conduction and valence bands respectively, with $\epsilon_c$ and $\epsilon_v$ being their respective energies at the wavevector $\mathbf{k}$. $\mathbf{e}_\alpha$ is the unit vector in Cartesian direction $\alpha$. To get the real part of the dielectric function we use the Kramers-Kronig transformation

\begin{equation}
    \varepsilon^{(1)}_{\alpha\beta}(\omega) = 1 + \frac{2}{\pi} \mathcal{P} \int_0^\infty \frac{\omega' \varepsilon^{(2)}_{\alpha\beta}(\omega')}{\omega'^2 - \omega^2+i\eta} d\omega',
\end{equation}

Where $\mathcal{P}$ is the principal value and $\eta$ is a Lorentzian broadening. For both the self-consistent DFT calculation to obtain converged wave functions and evaluation of the dielectric function, we use a 17$\times$17$\times$5 Gamma-centered mesh, and we use 540 bands to ensure that we have sufficient unoccupied bands for convergence of the summation. We select a Lorentzian broadening of 0.1 eV.

\subsection*{Crystal Growth, Sample Preparation and Thickness Measurements}

\MBT bulk crystals were grown using a Bi-Te flux, following the previously reported recipe in Ref.~\cite{yan_crystal_2019}. The magnetization data  in fields up to $\unit[12]{T}$ were collected using the AC option of a Quantum Design physical property measurement system.

In this work, we study thin bulk samples (``infinite'' layer) with a thickness of 136~nm, corresponding to approximately 100 septuple layers. These samples are similar to bulk crystals and are not air-sensitive, in contrast to atomically thin flakes which can degrade in air over time~\cite{otrokov_prediction_2019, ovchinnikov_intertwined_2021}. Nevertheless, care has been taken to minimize sample exposure to air before measurements.

To obtain thin bulk samples with atomically flat surfaces suitable for RCD studies, we used Scotch tape exfoliation of bulk crystals onto silicon wafers covered with 285~nm SiO$_2$ (we used 3M Magic Scotch Tape). The silicon wafers were pre-treated with RF O$_2$ plasma (duration: 10 minutes; power: 80~W; O$_2$ gas flow: 10~SCCM) before exfoliation to increase adhesion of crystals and overall yield. The exfoliation process was performed entirely inside a glovebox filled with inert argon gas, maintaining O$_2$ and H$_2$O levels below 0.1~ppm.

Once thin bulk samples of suitable dimensions were identified using an optical microscope, they were transported for loading into the low-temperature cryostat with optical access using a sealed container filled with argon. Care was taken to minimize air exposure during cryostat loading; we estimate that the sample was exposed to air for a maximum of 5 min before measurements. Based on the careful studies of oxidation on \ce{MnBi2Te4}~\cite{mazza_surface-driven_2022}, it is possible that the surface layer oxidized. However, the oxidation is self-limiting to the surface atomic layer~\cite{mazza_surface-driven_2022}, and therefore cannot dominate our optical measurements with~\unit[30]{nm} penetration depth. The height of the studied thin bulk sample was measured after RCD measurements were completed using atomic force microscopy (Asylum Research Cypher S, tapping mode) in air.


\onecolumngrid
\newpage
\newpage
\section*{Extended data}

\counterwithout{equation}{section}
\renewcommand\theequation{ED\arabic{equation}}
\renewcommand\thefigure{ED\arabic{figure}}
\renewcommand\thetable{ED\arabic{table}}
\renewcommand\thesection{ED\arabic{section}}
\renewcommand\bibnumfmt[1]{[ED#1]}
\setcounter{equation}{0}
\setcounter{figure}{0}
\setcounter{enumiv}{0}

\begin{figure}[h]
    \centering
    \includegraphics[width=1\linewidth]{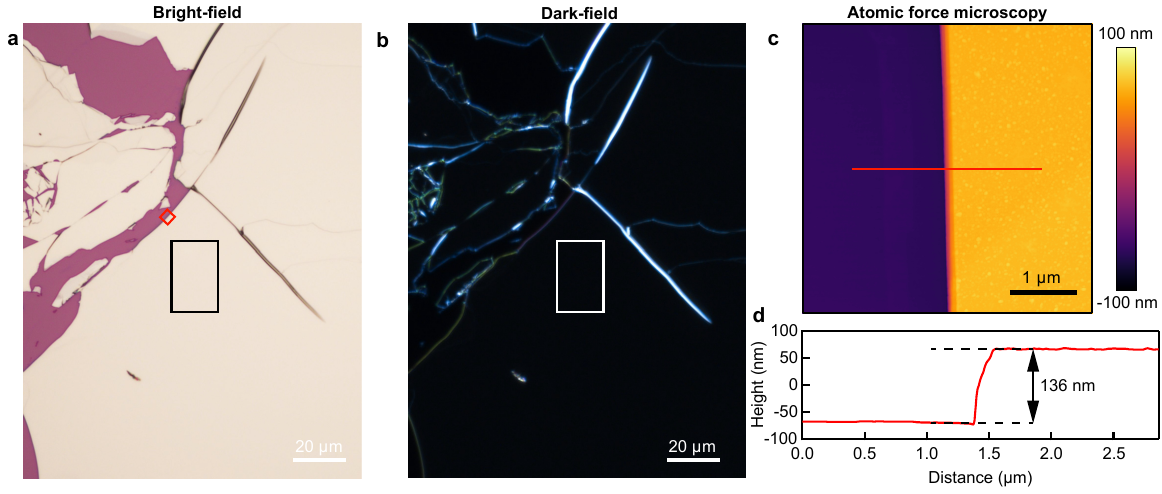}
    \caption{
         \textbf{Optical and atomic force microscopy characterization of thin bulk \MBT.}
        (a) Optical micrograph (bright field) of the region of interest. The black rectangle outlines the area where RCD sweeps and maps were taken. 
        The red rectangle shows the area where the height was measured using atomic force microscopy. 
        (b) Optical micrograph (dark field) of the same region with a white rectangle demonstrating the area where RCD was measured. 
        (c) Atomic force microscopy image of the edge of the MnBi$_2$Te$_4$ thin bulk flake used for optical measurements (outlined in the red rectangle in panel (a)). (d) The height profile measured along the red line in (c) which measures the height of the thin bulk to be 136~nm.}\label{fig:SamplePhoto}
\end{figure}

\newpage
\begin{figure}[p]
    \centering
    \includegraphics[width=1 \linewidth]{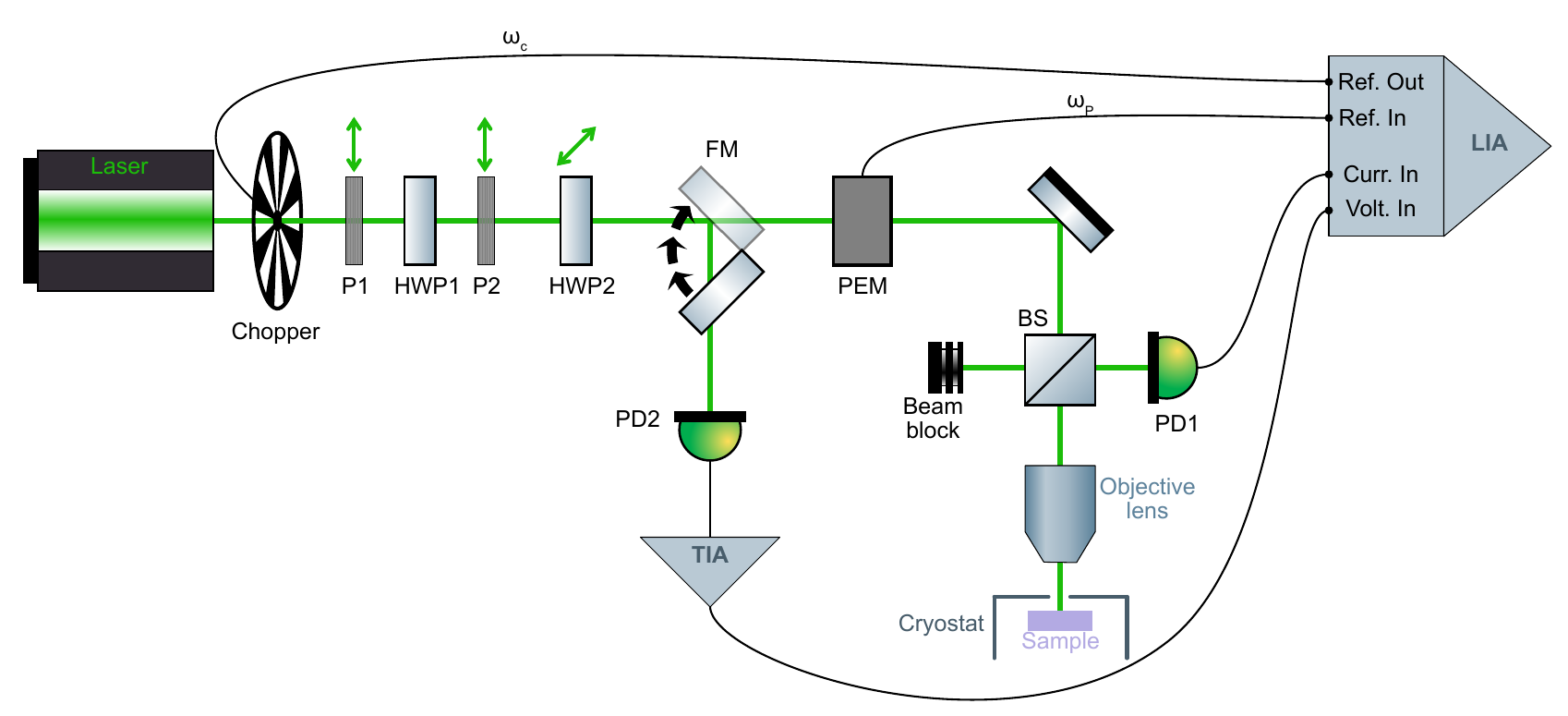}
     \caption{
        \textbf{Optical setup used for reflection circular dichroism (RCD) measurements.} The laser beam is modulated by a mechanical chopper at frequency $\omega_c$, and directed toward the sample via a series of optics which control light intensity and polarization. The first half-wave plate (HWP1), placed between vertical polarizers P1 and P2, is used to control the intensity. HWP2 sets the polarization angle to $45^\circ$ with respect to the modulation axis of the photoelastic modulator (PEM), which modulates the helicity of the beam at $\omega_P=\unit[50]{kHz}$. A flip mirror (FM) directs the light towards photodiode PD2 before the measurement at each wavelength, to ensure that the same intensity reaches the sample at each wavelength. The beam is focused onto the sample by a 10x objective (NA=0.25) that is outside of the cryostat. The reflected light is split by a beam splitter (BS), and focused onto photdiode PD1. The signals are read through the Zurich Instruments MFLI lock-in amplifier with two demodulators, at frequencies $\omega_c$ and $\omega_P$. The current from PD1 is read through the low-noise current input of the LIA, while the current from PD2 is transformed into voltage by a transimpedance amplifier (TIA), and read through the voltage input of the same LIA. 
        }
    \label{fig:setup}
\end{figure}

\newpage
\begin{figure}[p]
    \centering
    \includegraphics[width=0.7\linewidth]{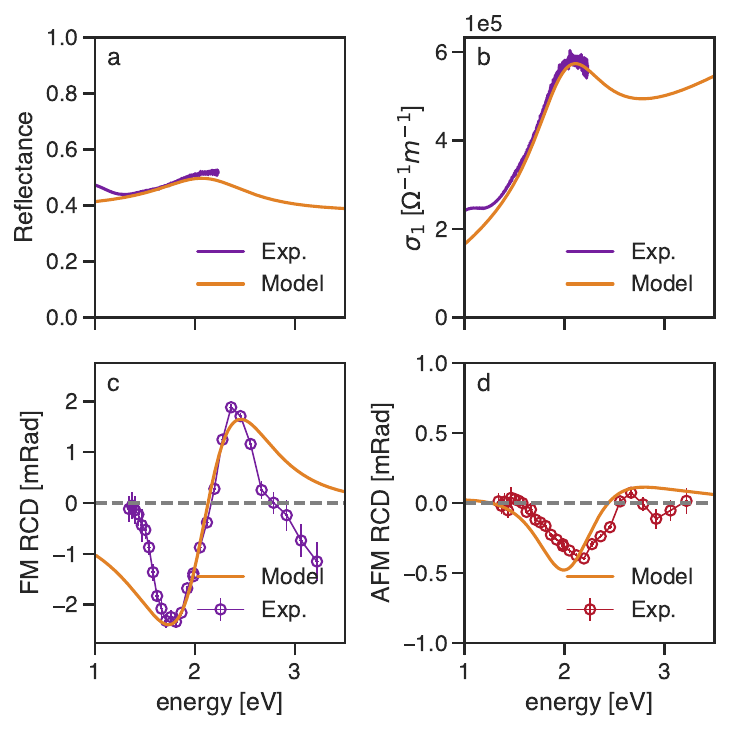}
    \caption{\textbf{Comparison of the experimental data to the Lorentz model}. (a) Reflectance ($R$), (b) the real part of conductivity ($\sigma_1$) and RCD of a (c) ferromagnet and (d) antiferromagnet calculated using the dielectric function described by the Lorentz model (Eq.~\ref{eq:Lorentz}, parameters listed in Table~\ref{tab:optical_params}), compared to experimental values. The experimental data for $R$ and $\sigma_1$ were published in Ref.~\cite{kopf_influence_2020}, while RCD is measured as a part of this work. The FM RCD curve from the model is multiplied by 0.4, since the measured jump in RCD corresponds to 0.4 of the saturation magnetization value at this temperature. The AFM RCD is calculated using the transfer matrix approach, assuming that each layer shares the dielectric properties of the bulk FM, with alternating spin orientation.}
    \label{fig:LorentzExpComparison}
    
\end{figure}

\begin{table}[p]
    \centering
      \renewcommand{\arraystretch}{1.2} 
    \setlength{\tabcolsep}{4pt} 
    \begin{tabular}{| c|  c|  c|  c | c|}
        \hline
         $\gamma / \omega_0$ & $ f / (\omega_0^2\varepsilon)$ & $\omega_0 [eV]$ & $\delta \omega [eV]$ & $\varepsilon_{\infty}/\varepsilon$ \\
        \hline
     
         0.424 & 4.2 & 2.05 & -0.0708 & 8.25 + 10.9 $i$ \\
    
        \hline
    \end{tabular}
    \caption{Parameters of the Lorentz model (Eq.~\ref{eq:Lorentz}), used for the figures in the main text (Fig.~\ref{fig:Fig3}b and Fig.~\ref{fig:Fig4}). 
    }
    \label{tab:optical_params}
\end{table}

\newpage
\begin{figure}[p]
    \centering
    \includegraphics[width=0.7
    \linewidth]{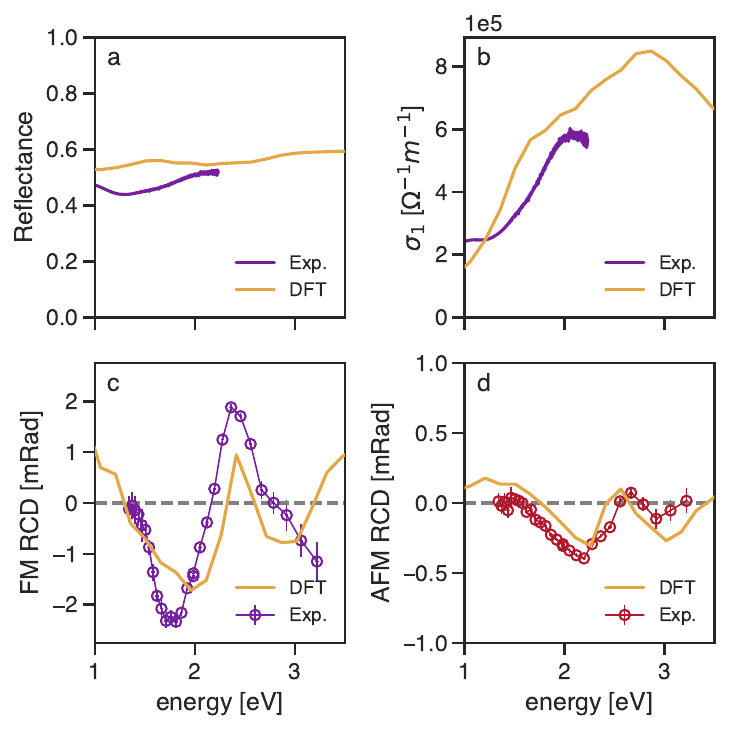}
    \caption{\textbf{Comparison of the experimental data to the quantities calculated from the DFT calculation.} (a) Reflectance ($R$), (b) the real part of conductivity ($\sigma_1$) and RCD of (c) a ferromagnet and (d) an antiferromagnet calculated using density functional theory, compared to experimental values. The experimental data for $R$ and $\sigma_1$ were published in Ref.~\cite{kopf_influence_2020}, while RCD is measured as a part of this work. 
    The DFT calculation is done for the polarized FM ground state, and the ordered moment is found to be equal to $\unit[4.548 ]{\mu_B/\text{Mn}}$. To directly compare with the experimental data we introduce the following scaling: the DFT FM RCD is multiplied by 1.2/4.548, reflecting the magnitude of the magnetization change at the spin-flop transition at $\unit[15]{K}$, while the DFT AFM RCD curve is multiplied by 3/4.548, reflecting the fact that the ordered moment at $\unit[15]{K}$ is $\unit[3]{\mu_B/\text{Mn}}$. The AFM RCD is calculated using the transfer matrix approach, assuming that each layer shares the dielectric properties of the bulk FM, with alternating spin orientation. The DFT data in all four panels were rigidly shifted by $\unit[300]{meV}$.}
    \label{fig:DFTModels_offset}
\end{figure}

\begin{figure}[p]
    \centering
    \includegraphics[width=0.90\linewidth]{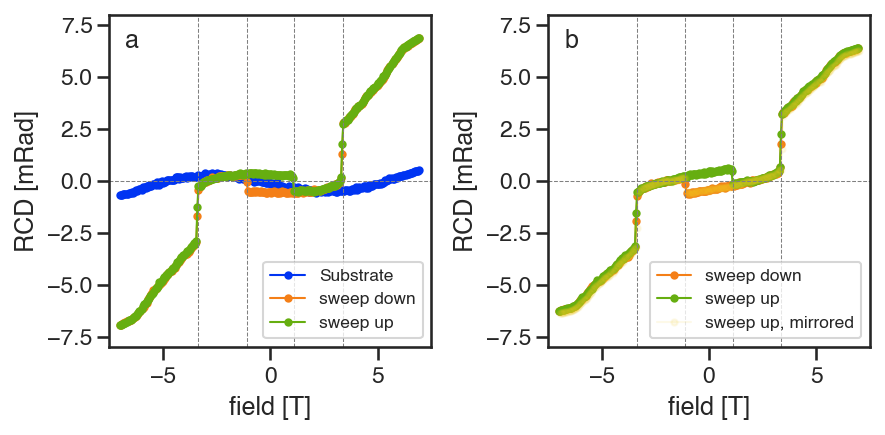}
    \caption{\textbf{RCD background subtraction} (a) RCD measured on a non-magnetic Si/SiO$_2$ substrate, as well as RCD measured on \MBT~ on upward and downward field sweeps. (b) The measured signal with the background subtracted, also shown in Fig.~\ref{fig:Fig2}a. We include the mirrored version of the `sweep up' curve, which perfectly aligns with the `sweep down' curve. This indicates that the two states at $H = 0$ are time-reversed versions of each other, and is inconsistent with the hypothesis that only a top surface ferromagnetic spin is reoriented by the magnetic field.}
    \label{fig:BackgroundSubtraction}
\end{figure}

\begin{figure}[p]
    \centering
    \includegraphics[width=0.85\linewidth]{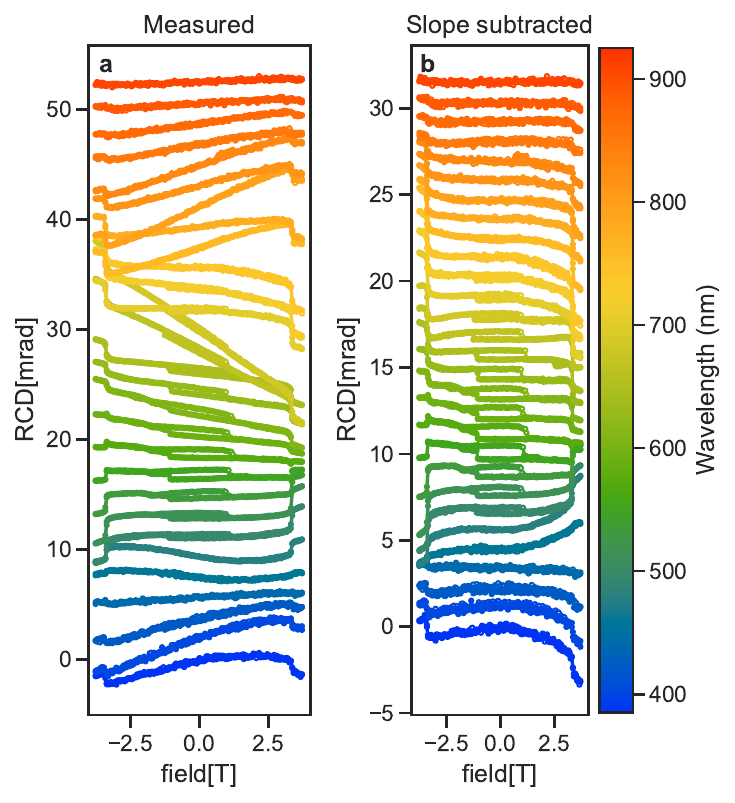}
    \caption{\textbf{RCD \textit{vs} field for the range of wavelengths used in Fig.~\ref{fig:Fig3}a.} (a) The measured RCD vs magnetic field, measured at a range of wavelengths. (b) Same as (a), but with the slope at $H=0$ subtracted for clarity. Setup artifacts contribute a field- and wavelength- dependent background to these measurements, however they do not impact the extracted values of RCD at $H=0$ or across the spin-flop transition, because both are obtained through differential measurements, that cancel the background effects (see Fig.~\ref{fig:Fig3}a, and the corresponding caption). Curves are offset for clarity. }
    \label{fig:WavelengthFieldDependence}
\end{figure}

\begin{figure}[p]
    \centering
    \includegraphics[width=0.95\linewidth]{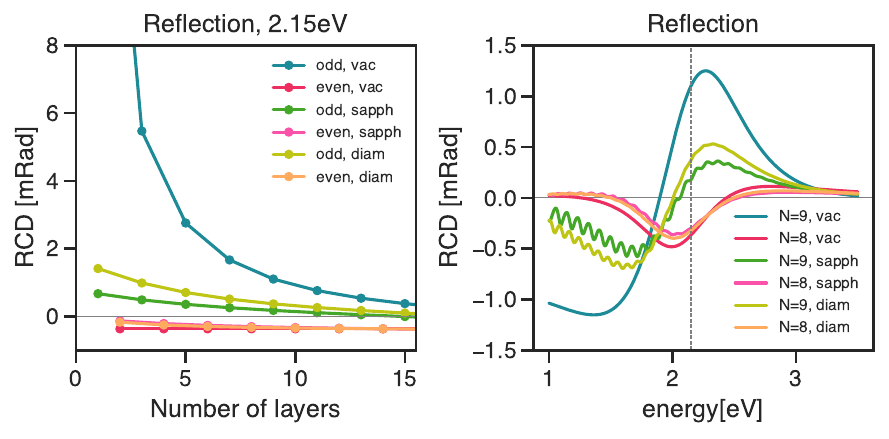}
    \caption{\textbf{Comparison of RCD for samples suspended in vacuum and on two substrates, sapphire and diamond.} (a) RCD as a function of layer number, for photon energy of $\unit[2.15]{eV}$. Odd-layer RCD is significantly suppressed for samples on substrates compared to the ones suspended in vacuum. (b) RCD spectra for $N=8$ and $N=9$, showing that samples on substrates exhibit smaller differences in RCD magnitude for even and odd layer flakes across the spectrum. Calculation is done using the Lorentz model (Eq.~\ref{eq:Lorentz}), with parameters listed in Table~\ref{tab:optical_params}. The sapphire substrate is modeled with an energy independent dielectric constant $\varepsilon=\varepsilon_0 \left(10+10^{-4}i\right)$, and the diamond with $\varepsilon=\varepsilon_0 \left(5.7+10^{-4}i\right)$, where we added the small imaginary components to suppress multiple reflections, which give rise to the oscillatory component of the RCD \textit{v.s.} energy. The thickness of the substrates is taken to be $\unit[1]{cm}$.}\label{fig:Sapphire}
\end{figure}

\clearpage

\counterwithout{equation}{section}
\renewcommand\theequation{S\arabic{equation}}
\renewcommand\thefigure{S\arabic{figure}}
\renewcommand\thetable{S\arabic{table}}
\renewcommand\thesection{S\arabic{section}}
\renewcommand\bibnumfmt[1]{[S#1]}
\setcounter{equation}{0}
\setcounter{figure}{0}
\setcounter{enumiv}{0}

\newpage
\setcounter{linenumber}{1}
\setcounter{page}{1}    
\section*{Supplementary Information for `Universal Kerr effect in an A-type antiferromagnet'}
\subsection*{The analytical solutions}

\subsubsection*{Bulk  ferromagnet}

The reflectivity matrix of a ferromagnet in the $\mathbf{\hat{e}}_{\pm}$ basis is: 

\begin{align}
    r^{FM}_C &=\begin{pmatrix}
        r_0+\delta r & 0\\
        0 & r_0 - \delta r
    \end{pmatrix}=
    \begin{pmatrix}
        \frac{1-n-\delta n}{1+n+\delta n} & 0\\
        0 & \frac{1-n+\delta n}{1+n-\delta n},
    \end{pmatrix}.
\end{align}
where $n_{1,2}=n\pm\delta n$ are the indices of refraction for $\mathbf{\hat{e}}_{+}$ and $\mathbf{\hat{e}}_{-}$,  that is for LCP and RCP incident light, respectively. This leads to: 

\begin{equation}
    \delta r = \frac{1}{2} \left(\frac{1-n-\delta n}{1+n+\delta n}-\frac{1-n+\delta n}{1+n-\delta n}\right)=\frac{-2 \delta n}{-\delta n^2+n^2+2 n+1}
\end{equation}
and
\begin{equation}
    r_0 = \frac{\delta n^2-n^2+1}{-\delta n^2+n^2+2 n+1},
\end{equation}
yielding the complex Kerr angle (Eq.~\ref{eq:KerrDef}):
\begin{equation}
    \Theta_{FM}= (-i) \left(-\frac{2 \delta n}{\delta n^2 - n^2 + 1}\right)\approx \frac{-2i\delta n}{n^2-1}.
\end{equation}

\subsubsection*{Antiferromagnetic bilayer}
We use the transfer matrix method to analytically solve for the reflection and transmission of an antiferromagnetic bilayer suspended in vacuum (Fig.~\ref{fig:Fig5}a). The indices of refraction for $\mathbf{\hat{e}}_{+}$ polarization of the two layers $n_{1,2}=n\pm\delta n$, and $n_{1,2}=n\mp\delta n$ for $\mathbf{\hat{e}}_{-}$. The thickness of the individual layers is $d$, and $k_0=\omega/c$, were $\omega$ and $c$ are the frequency and speed of light. 

We formulate the calculation in the basis of circular polarization, so that the problem is diagonal. We find that the transmission circular dichroism vanishes identically. In contrast, for reflectivity we find:

\begin{equation}
    \frac{\delta r}{r_0} = \frac{2i \, \delta n \, n \left(\cos \left(2dk_0 n\right) - \cos \left(2dk_0 \delta n\right)\right)}
    {n (-\delta n^2 + n^2 - 1) \sin \left(2dk_0 n\right) 
    + \delta n (-\delta n^2 + n^2 + 1) \sin \left(2dk_0 \delta n\right)}
\end{equation}

The complex MOKE angle can be found using Eq~\ref{eq:KerrDef}. In the physically relevant limit of $dk_0\ll1$, $\delta n\ll n$, this expression reduces to 

\begin{equation}
    \Theta_{AFM} = (-i)\frac{\delta r}{r_0}\approx - \frac{2    \delta n}{n^2-1}(dk_0n)=- \frac{2i\delta n}{n^2-1}(-idk_0n)=\Theta_{FM}(-idk_0n)
\end{equation}
where $\Theta_{FM}$ stands for the complex Kerr angle of a bulk ferromagnet described by the $n \pm \delta n$ for $\mathbf{\hat{e}}_{+}$. 

\subsection*{Photoelastic modulator and the RCD measurement}

The time-dependent effect of the photoelastic modulator (PEM) can be captured in the basis of linear polarization with the following Jones matrix: 

\begin{equation}\label{eq:PEM}
    J_{PEM}=\left(
\begin{array}{cc}
 e^{-\frac{1}{4} i \pi  \sin (t \omega_P )} & 0 \\
 0 & e^{\frac{1}{4} i \pi  \sin (t \omega_P )} \\
\end{array}
\right),
\end{equation}
where $\omega_P$ is the angular frequency of polarization modulation. The modulated light is reflected from the sample, and focused onto a photodiode. The modulation of total reflectivity that is synchronous with the polarization modulation is proportional to the imaginary part of $\Theta$, which can be seen from a Jones matrix calculation, outlined below.

The reflected electric field is given by,

\begin{equation}\label{eq:outputField}
    \begin{pmatrix}
        E_x \\
        E_y
    \end{pmatrix}= r_L\times J_{PEM}\times \frac{1}{\sqrt{2}}\begin{pmatrix}
        1 \\
        1
    \end{pmatrix},
\end{equation}
where $r_L$ is the sample reflectivity expressed in the linear basis. 
The measured quantity is intensity,

\begin{equation}
    I=\left|E\right|^2 \left[\frac{1}{2}\left(1+ \operatorname{sgn}\left(\sin{(\omega_c t)} \right)\right)\right],
\end{equation}
where the square wave term in the square brackets accounts for the modulation by the mechanical chopper at frequency $\omega_c$, and $\left|E\right|$ is the absolute value of the reflected electric field. Keeping only the first term in the Fourier transform of the square wave, we have: 

\begin{equation}
    I=\left|E\right|^2 \left[\frac{1}{2}+ \frac{2}{\pi}\sin{(\omega_c t)}\right].
\end{equation}
To calculate $\left|E\right|^2$, we use Eqs.~\ref{eq:Rlinear},~\ref{eq:PEM} and~\ref{eq:outputField}, and find:
\begin{equation}
        \frac{\left|E\right|^2}{\left|r_{0}\right|^2}= 1+\left|\frac{r_{xy}}{r_0}\right|^2+2 \operatorname{Im}\left(\frac{r_{xy}}{r_{0}}\right) \sin \left(\frac{1}{2} \pi  \sin \left(\omega_P t\right)\right) \approx 1+4J_1\left(\frac{\pi}{2}\right) \operatorname{Im}\left(\Theta\right)  \sin \left(\omega_P t\right),
\end{equation}

\noindent where $J_1\left(\pi/2\right)$ is the Bessel function of the first kind. In the last step we used the Jacobi-Anger expansion, and omitted the quadratic term in $r_{xy}/r_{0}$. The key point is that this signal contains information on $\operatorname{Im}\left(\Theta\right)$. 

The intensity can  be expressed as:
\begin{align}
    &\frac{I}{\left|r_{0}\right|^2} =\left(1+4J_1\left(\frac{\pi}{2}\right) \operatorname{Im}\left(\Theta\right)  \sin \left(\omega_P t\right)\right) \left[\frac{1}{2}+ \frac{2}{\pi}\sin{(\omega_c t)}\right]\\
    &=\frac{1}{2}+ \frac{2}{\pi}\sin{(\omega_c t)}+2J_1\left(\frac{\pi}{2}\right) \operatorname{Im}\left(\Theta\right)  \sin \left(\omega_P t\right)+\frac{8}{\pi}J_1\left(\frac{\pi}{2}\right) \operatorname{Im}\left(\Theta\right)  \sin \left(\omega_P t\right)\sin{(\omega_c t)}.
\end{align}

We simultaneously demodulate the measured intensity at $\omega_c$ and $\omega_P$ using a lock-in amplifier, yielding currents $I_c$ and $I_P$, respectively. We find: 
\begin{equation}
    \frac{I_P}{I_c}=\pi J_1\left(\frac{\pi}{2}\right) \operatorname{Im}\left(\Theta\right),
\end{equation}
or, equivalently,

\begin{equation}
    \operatorname{Im}\left(\Theta\right)= \frac{I_P}{I_c}\frac{1}{\pi J_1\left(\pi/2\right)}\approx\frac{I_P}{I_c}\frac{1}{1.78073}.
\end{equation}

\section*{Effect of Surface Quality on Coercive Field and Zero-Field RCD}

\begin{figure}[h]
    \centering
    \includegraphics[width=0.50\linewidth]{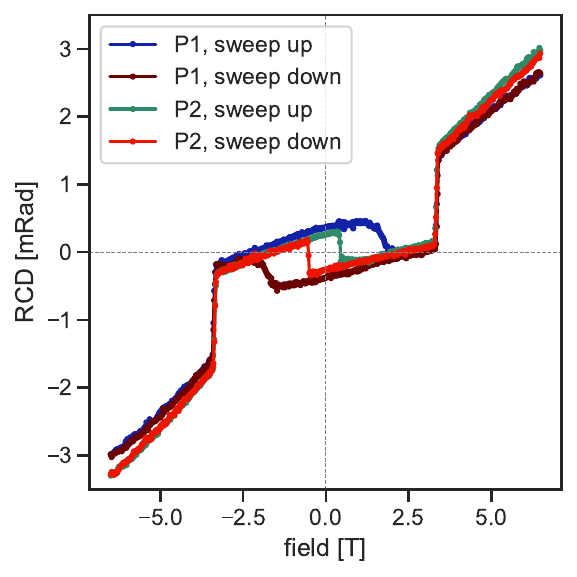}
    \caption{\textbf{Flake-to-flake variation.} RCD as a function of field for two different bulk \ce{MnBi2Te4} flakes, measured at a wavelength of 570\,nm. While the coercive field varies between flakes, the zero-field RCD magnitude is identical within experimental error. This demonstrates that the coercive field is sensitive to surface-specific details, whereas the zero-field RCD is a robust property of the bulk antiferromagnetic state.}
    
    \label{fig:Two_flakes}
\end{figure}

The coercive field in our measurements varies between nominally identical flakes exfoliated from the same MnBi$_2$Te$_4$ crystal, whereas the magnitude of the reflection circular dichroism (RCD) at $H = 0$ remains constant. This observation is illustrated in Fig.~\ref{fig:Two_flakes}, in which we compare two flakes placed on the same substrate and measured in the same cooldown. While the zero-field RCD values are identical within experimental error, the coercive fields and the shapes of the hysteresis differ.

We attribute this difference in coercivity to variations in local surface conditions, such as strain introduced during exfoliation or subtle differences in surface oxidation. We know that the Hamiltonian parameters of the surface spin (exchange coupling, anisotropy) are different from those of the bulk, and as we point in the main text this difference is crucial for magnetic-field tuning of AFM domains. The difference between the Hamiltonian properties at the surface and the bulk can of course be influenced by details of the surface, leading to the different coercivity. However, surface oxidation in MnBi$_2$Te$_4$ is self-limiting and confined to the topmost atomic layer~\cite{mazza_surface-driven_2022}, and the optical penetration depth far exceeds this layer thickness. Consequently, while the surface environment can affect the switching field of the antiferromagnetic domains, it cannot account for the robust and reproducible RCD magnitude at $H = 0$.

To further test whether surface modifications could mimic our results, we simulated RCD spectra for three scenarios: (a) pristine A-type AFM with domain reversal; (b) A-type AFM with domain reversal, with a non-magnetic surface layer; (c) AFM with a ferromagnetic surface layer that switches independently of the bulk. Only scenarios (a,b), in which the AFM domain switches, reproduce the experimental finding that the two zero-field RCD magnitudes are equal in magnitude and opposite in sign (Fig.~\ref{fig:BackgroundSubtraction}). In particular, switching only a surface ferromagnetic layer yields asymmetric zero-field spectra, inconsistent with the data (Fig.~S2).

\begin{figure}
    \centering
    \includegraphics[width=1\linewidth]{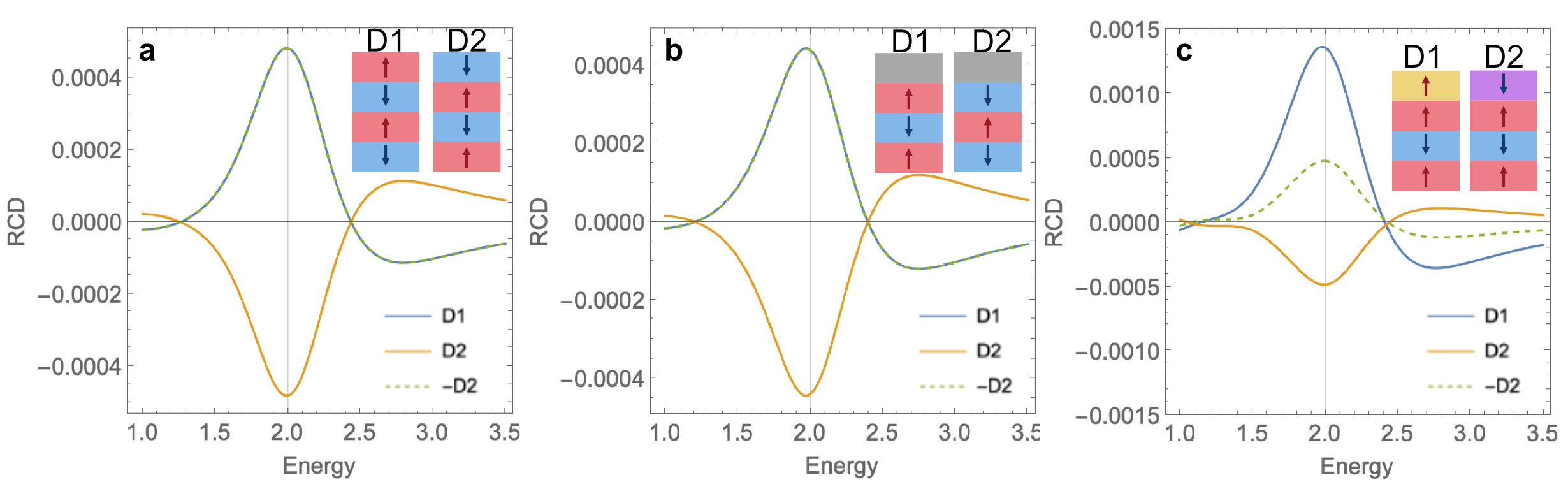}
\caption{\textbf{Effect of surface modifications on zero-field RCD.} Simulated RCD spectra for three scenarios: (a) pristine A-type AFM with domain reversal; (b) A-type AFM with domain reversal and a non-magnetic surface layer; and (c) A-type AFM with a ferromagnetic surface layer that switches independently of the bulk. Each panel shows RCD as a function of photon energy for the two AFM domain configurations sketched in the corner. Only scenarios (a) and (b), in which the AFM domain switches, reproduce the experimental observation that the two zero-field RCD spectra are equal in magnitude and opposite in sign. Switching only a surface ferromagnetic layer yields asymmetric zero-field spectra, inconsistent with the experiment.}

 \label{fig:placeholder}
\end{figure}
These findings establish that the reproducible zero-field RCD signal arises from reversal of bulk AFM domains. While the coercive field is sensitive to details of the surface, this does not influence the magnitude of the RCD at $H = 0$, which is a robust signature of the underlying bulk antiferromagnetic order.

\end{document}